\documentclass{article}
\usepackage{misc/arxiv3}

\usepackage{amsmath}
\usepackage{url} 
\usepackage{amsthm} 
\usepackage[colorlinks=true,linkcolor=red,filecolor=green,citecolor=blue]{hyperref}
\usepackage{comment}
\usepackage{graphicx} %for dealing with figures.
\usepackage{caption}
\usepackage{subcaption}
\usepackage{algorithmic} %for describing algorithms
\usepackage{url}% It provides better support for handling and breaking URLs.
\PassOptionsToPackage{hyphens}{url}
\usepackage{hyperref}
\expandafter\def\expandafter\UrlBreaks\expandafter{\UrlBreaks\do\-\do\~\do\'\do\"\do\-}%

\usepackage{changepage}
\usepackage{appendix}

\usepackage[dvipsnames]{xcolor} 

\definecolor{colour1}{RGB}{166,206,227}
\definecolor{colour2}{RGB}{31,120,180}
\definecolor{colour3}{RGB}{178,55,250} % purple
\definecolor{colour4}{RGB}{51,160,44}

\newcounter{noteMCctr} 

\newcommand{\mc}[1]{{\textcolor{black}{{{}}#1}}{}}%colour3

\newcounter{noteGRctr} 

 \newcommand{\gr}[1]{{\textcolor{black}{{{}}#1}}{}}%magenta

 \newcommand{\gdr}[1]{{\textcolor{black}{{{}}#1}}{}}%magenta
\newcounter{noteAMctr} 

 \newcommand{\am}[1]{{\textcolor{black}{{{}}#1}}{}}%yellow

\newcommand{\kh}[1]{{\color{black} #1}}
\RequirePackage[authoryear]{natbib}
\title{GDP nowcasting with large-scale inter-industry payment data in real time -- A network approach}

\author[1]{Anastasia Mantziou}
\author[2]{Kerstin H\"otte}
\author[3,4]{Mihai Cucuringu}
\author[3,4]{Gesine Reinert}
\affil[1]{Department of Statistics, University of Warwick, Coventry, CV4 7AL, UK}
\affil[2]{Department of Accounting, Economics \& Finance, KEDGE Business School, 40 Avenue de Terroir de France, Paris 75012, France}
\affil[3]{Department of Statistics, University of Oxford, 24-29 St Giles, Oxford OX1 3LB, UK}
\affil[4]{The  Alan Turing Institute, British Library, 96 Euston Rd., London, NW1 2DB,  UK}

\date{}

\begin{document}

\maketitle

\begin{abstract}
Real-time economic information is essential for policy-making but difficult to obtain. We introduce a granular nowcasting
method for macro- and industry-level GDP using a network approach and data on real-time monthly inter-industry
payments in the UK. To this purpose we devise a model which we call an extended generalised network autoregressive
(GNAR-ex) model, tailored for networks with time-varying edge weights and nodal time series, that exploits the notion
of neighbouring nodes and neighbouring edges. The performance of the model is illustrated on a range of synthetic data
experiments. We implement the GNAR-ex model on the payments network including time series information of GDP
and payment amounts. To obtain robustness against statistical revisions, we optimise the model over 9 quarterly releases
of GDP data from the UK Office for National Statistics. Our GNAR-ex model can outperform baseline autoregressive
benchmark models, leading to a reduced forecasting error. This work helps to obtain timely GDP estimates at the
aggregate and industry level derived from alternative data sources compared to existing, mostly survey-based, methods.
Thus, this paper contributes both, a novel model for networks with nodal time series and time-varying edge weights, and
the first network-based approach for GDP nowcasting based on payments data.
\end{abstract}


%{   \hypersetup{linkcolor=blue}
%   \tableofcontents
%}

\section{Introduction}

Real-time economic information is essential for policymaking in a rapidly changing and interconnected world, faced  
\gr{with} local and global shocks, technological change, and increasing environmental and social pressure. Yet, such information is difficult to obtain.  
\gr{In particular,} GDP is a major indicator used in decision-making, but its compilation is complex and draws on hundreds of data sources, which leads to long publication delays. Autoregressive estimates fail to capture unanticipated events, leading to hefty revisions when the flow of incoming information reveals an increasingly detailed picture of the state of the economy \citep{OSR2023revisions, barrand2021meeting}. 

Economic nowcasting can fill the data gap, by using alternative, observable real-time information to predict current economic outcomes. It has attracted much high-level policy interest, particularly in the aftermath of COVID-19 and recent energy price shocks \citep{woloszko2020weekly, ons2023realtime}. Examples of nowcasting include methods relying on aggregate real-time information extracted from Economic News \citep{barbaglia2023forecasting}, Google Data \citep{ferrara2023google, woloszko2020weekly}, seismographic patterns caused by industrial activities \citep{tiozzo2022seismonomics}, payments data \citep{galbraith2018nowcasting, esteves2009atm, carlsen2010dankort, das2021gdp, aprigliano2019using, duarte2017mixed}, \am{or utilising various economic indicators \citep{koop2024incorporating,proietti2021nowcasting}}. 

Most of these approaches rely on aggregate data and use mixed frequency sampling or factor models to extract statistical signals from various time series that are characterised by high collinearity and asynchronous updating and availability \citep{banbura2013now, bok2018macroeconomic}.  
Such aggregate methodology hides the mechanisms and channels of how observed aggregate phenomena drive economic fluctuations.
Understanding these channels is relevant for targeted policies aimed at preventing and mitigating shocks, that propagate through an interconnected economy, and at responding to their effects at the micro and macro level on time \citep{acemoglu2012network, whitehouse2023supply}. 

Supply chains are a major propagation mechanism of economic dynamics and disruptions \citep{acemoglu2012network}, as powerfully illustrated by  
recent supply chain disruptions associated with COVID-19, the war in Ukraine, and natural hazards. \gr{This observation suggests pursuing ideas from network analysis for GDP nowcasting.} 
\am{\gr{Most n}etwork autoregressive models however 
focus 
on modelling time series \gr{which are} observed on the nodes of a fixed network; } \gr{here is a brief survey.} \cite{zhu} introduce a network autoregressive model tailored for networks with nodal time series, which includes a network effect parameter that explains the impact of neighbouring nodes  
for each node 
in the network. \cite{gnar} generalise the model by \cite{zhu} by using the notion of higher-stages of node neighbours which also allows to capture the impact of  
time series \gr{between nodes that are not direct neighbours in the network.} 
\cite{martin2024nirvar}  
introduce a network autoregressive model for panels of multivariate time series represented through a multiplex network, assuming that the network is weighted with fixed weights, and \gr{has a stochastic block model} 
(SBM) structure. 
\gr{While these three models} assume Gaussian noise,  
\cite{armillotta2024count} propose a network autoregressive model for count time series observed on the nodes of a fixed network \gr{with} 
Poisson \gr{distributed} responses.

\gr{In contrast, the payment data between industry groups are more naturally represented as edge weights in a network. For time series on the edges of a network,} in  \cite{mantziou2023gnar} it is shown that a network approach with nodes representing industries and weighted edges representing transactions between industries captures subtle dependencies in the data which a standard vector autoregressive (VAR) model would not include.
 \gr{However time series on nodes, which could for example model GDP, are not included in the model by \cite{mantziou2023gnar}.}

In this paper, we introduce a theoretically informed network-based approach to GDP nowcasting at the industry and macroeconomic level, by using novel \gdr{experimental} real-time payment data that captures supply chain dynamics among UK industries \citep{ons2023interindustry, hotte2024}\gr{, in combination with quarterly GDP estimates provided by the UK Office for National Statistics (ONS)}. 
We contribute a granular model to nowcast macro- and industry-level GDP \citep{ons2023indicativeGDPdata}  
\gr{which we coin}  
\textit{GNAR-ex},  
\gr{extending current} 
network autoregressive models  \citep{gnar,mantziou2023gnar}. Nodes in the network represent industries, and edges are payment flows between them. 
\gr{On} the \gr{(fixed)} network \gr{with industry groups as nodes}, we observe two types of time series: time series of GDP on the nodes (industries) and time series of payment flows \gr{between industries, represented as} 
time-varying weights on the edges of the network. The model is a restricted VAR model \citep{lutkepohl2005new}, with GDP being estimated as a function of lagged GDP and lagged payment flows according to the network structure through the notions of neighbouring nodes and neighbouring edges.

Revisions are a major issue in economic nowcasting; GDP estimates for a given time stamp differ across data releases \citep{hepenstrick2019forecasting}. \gr{For the ONS data, initial estimates are revised in subsequent quarterly releases, until after around 36 months a final GDP figure is achieved.} 
We combine 9 \gr{subsequent}  
\gr{ONS} releases of GDP data to obtain 
\gr{a nowcasting method} that is robust 
\gr{across} revisions. 
\gdr{Here we stress that the results are based on experimental data.}

{\gr{Here is a summary of our} contributions.} 
\begin{enumerate}
    \item We introduce a novel model, the GNAR-ex model, that combines time series information observed both on the nodes
and the edges of a fixed network, using information about the network structure through combining the notions of
neighbouring nodes and neighbouring edges in the same network autoregressive model. We demonstrate the estimation
performance and predictive power of the GNAR-ex model in synthetic data experiments. Moreover, we address model
uncertainty by model averaging (MA) and explore the accuracy of the MA forecasts in synthetic data experiments.
\item We apply the GNAR-ex model on the real data to provide nowcasts of GDP at the industry and macroeconomic level.
The model parameters are estimated via least-squares, using data from a given release, and the model predicts GDP
of the next period as published in the subsequent ONS release; this nowcast is compared to the official first ONS
estimate of GDP for that quarter. We calculate the relative error and benchmark the predictive performance against
an ARIMA model which does not include network information.
\item To address multicollinearity, we apply network sparsification methods informed by economic intuition and time series
correlations. We observe that network sparsification significantly improves the predictive power.
\item Investigating the performance release-by-release, we show that the GNAR-ex model outperforms the best-fitting
ARIMA benchmark, but the optimal model configuration varies by the number of lags and network neighbours being
included. For real-life nowcasting, a more practical approach is required. To obtain a uniform and well-performing
model, we apply MA and show that the GNAR-ex model is competitive with an ARIMA(0,1,0), which was suggested
as one of the best performing models using the R function \texttt{auto.arima}.
\item To the best of our knowledge, our approach delivers the first statistical network time series model applied to economic
nowcasting of GDP. Economists have shown that the topology of supply chain networks can be decisive for explaining
aggregate fluctuations, with large industrial hubs and well-connected industries being key drivers of GDP volatility
\citep{acemoglu2012network}. Such models often reduce the network impact to a static vector of influence at the node
level, describing the impact of individual sectors on macroeconomic outcomes. Instead, the GNAR-ex model explicitly
includes fluctuations on the network edges in a structural model that combines node- and edge-level information. To
highlight the novelty of our approach, network models have rarely been used in economic nowcasting, albeit there
is related work; \citet{macqueen2023sectoral} use network information embodied in input-output tables to infer statistical
relationships between GDP components, to incorporate timely information when making predictions. Our approach
is different to \cite{macqueen2023sectoral} since our novel data set for payments data dictates the network structure and
allow inferences for network parameters explicitly.
\end{enumerate}

The remainder of this article is structured as follows: The next section (Sec.~\ref{sec:data}) explains the payment and GDP data. Sec.~\ref{sec:model} introduces the model and Sec.~\ref{sec:simulations} demonstrates its performance in synthetic data experiments.  In Sec.~\ref{sec:nowcastingGDP}, we show our main empirical results; Sec.~\ref{sec:conclusions} \gr{provides a discussion of the approach.}  Appendix \ref{sec:appendix} contains  additional results from synthetic data experiments and the real data application. 
 
\section{Data}\label{sec:data}
We combine two publicly available monthly time series data: (1) payment flows between industries, and (2) industry-level gross value added (GVA) data. \gr{On the node (industry) level, we observe time series of \kh{month-to-month} GDP {growth rates} $G_{i,t}$ for each industry $i$ at time $t$, while on the edges we observe time series of \kh{month-to-month} payment {growth rates} $P_{ij,t}$ between industries $i,j$ at time $t$.}
\gr{Our forecasting goal is}
the cross-industrial aggregate \gr{growth rate} of GVA, \gr{which we view} as a proxy of UK GDP. 

\subsection{The payment data}

The \gr{payment} data was published by the ONS in December 2023, as an experimental dataset with ongoing development. \gr{It is a data set which is an monthly \kh{industry-level}  aggregate of}
business-to-business (B2B) financial transactions made through the Bacs Payment System, which is one of the major routes of how businesses execute regular payments in the UK. Businesses were mapped to SIC codes and payments were aggregated by industry at a monthly basis, \gr{yielding 38 industry groups}. In principle, such data can be extracted in real-time from the payment infrastructure of the Bacs Payment System, making it a promising source of real-time economic indicators. Broadly speaking, the payments reflect trade in intermediates between industries. \citet{ons2023interindustrymethods, hotte2024} provide a more detailed description, validation and conceptual discussion of the data. 

\gr{We represent this} time series of inter-industrial payment data 
as a network time series \gr{on a simple directed network with $K$ industry groups as nodes, and} 
the flows of payments between a pair of  
\gr{industry groups} $i,j \in \gr{1, \ldots, K}$ \gr{as edge weights}; \gr{self-loops are excluded.}
While the edge weights change monthly, the set\kh{s} of nodes \kh{and} \am{edges are} assumed to be fixed. Thus, we construct a static \mc{directed} network with nodes representing the industries
and edges \gr{between two industries} encoding that payments between the two industries, made across time, could be of interest. \gr{This network is constructed before the data are inspected.}  
As the above formulation indicates, not all payments between industries may be of interest. Moreover, in a given month there may be some industry pairs for which no payments are observed, but a corresponding edge in the static network is still present. 
Details can be found in Section \ref{sec:nowcastingGDP}.

\subsection{Gross value added}

The Gross Value Added (GVA) data is obtained from the \citet{ons2023indicativeGDPdata}. It represents a series of quarterly data releases of monthly industry-level GVA index data, \gr{which is} consistent with quarterly national accounts and distinguishes 99 distinct industries, classified by Classification of Products by Activity (CPA) codes. To harmonise the aggregation with the 38 industries available in the payment data, we transformed the index data (2019=100) into levels of  
\gr{GVA} using industry-level GVA data from the 2019 Supply-and-Use Tables (SUT) \citep{ons2023blue} to obtain appropriate industry-weights before aggregating multiple CPAs into broader industrial categories.

One major issue in nowcasting is the continuous updating of early estimates \citep{banbura2013now}, when new data becomes available. Hence, the first estimates of published GVA close to the publication date may be revised in future releases. Further, methods for data collection and compilation are subject to minor changes, mostly arising from methodological improvements. 
As a consequence, the monthly level of the GVA index of industry $i$ \mc{at a} given $t$ may differ across data releases. The effect of incoming data affects the most recent GVA estimates, while methodological revisions affect the whole time series, which starts in 1997. 

The analysis in this paper is based on nine different industry-level GVA data releases, being published on a quarterly basis between December 2021 and December 2023. This was the maximum number of available releases when downloading the data in May 2024. We excluded the March 2024 release, as our approach assumes to use only data that had been available in December 2023, which also coincides with the availability of payment data. \am{We use the March 2024 release only to evaluate the forecasts using the December 2023 release.}

In abuse of notation, for easier interpretability we abbreviate ``GVA as proxy for GDP'' by ``GDP''.
Similarly, instead of ``industry groups'' we use ``industry''.

\section{The model}\label{sec:model}

\begin{figure}
    \centering
    \includegraphics[scale=.4]{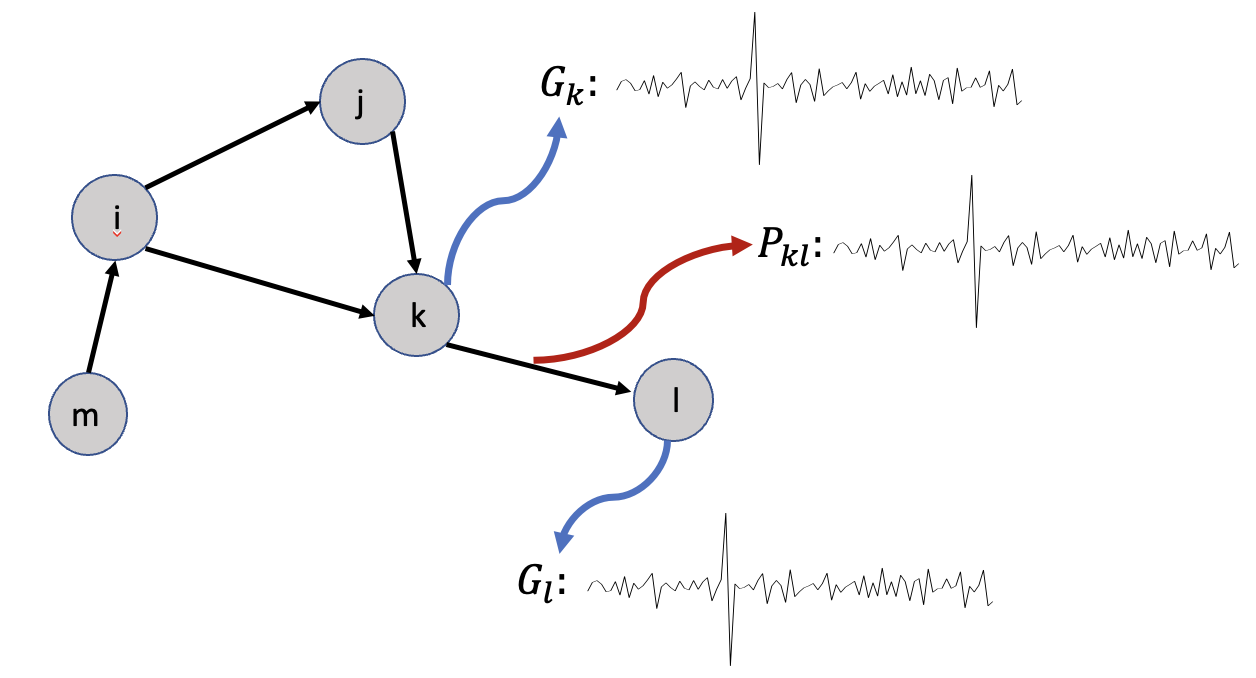}
    \caption{A toy example of a network with nodes representing industries and edges representing \gr{potential} transactions. Time series on the nodes represent time series of GDP {growth rates} and time series on the edges represent time series of transaction values growth rates.}
    \label{illnet}
\end{figure}

\mc{The model is anchored in the setting}
\kh{ of a}
\gr{
fixed network with $K$ nodes and $M$ edges,} \kh{and time series being} 
\gr{
%are
observed both on the nodes and on the edges.} 
{Figure \ref{illnet} shows a toy example
%.
}\kh{to illustrate the setting.}
%of the network.}
\gr{The GNAR-ex model is a time series model which starts with  a given static, directed, unweighted  network on $K$ nodes, without self-loops or multiple edges.}
As in \citet{gnar} and \citet{mantziou2023gnar}, 
{the} set of nodes is denoted by $ \mathcal{V}$  
and $\mathcal{E}$ denotes the set of 
edges\gr{; we assume that $|\mathcal{E}| =M$}. The set {$\mathcal{N}_{node}^{1}(i)$} of \textit{1-stage neighbouring nodes} for node $i$ is the set of all nodes for which a directed edge from or to node $i$ exists, defined as
$$\mathcal{N}_{node}^{1}(i)=\{j\in \mathcal{V}/\{i\}: \{i,j\}\in \mathcal{E} \text{ or } \{j,i\}\in \mathcal{E}\}$$
{F}or $r 
\ge 2$, we  
{denote} the set of $r
$-stage neighbouring nodes of node $i$ 
as the set 
$$\mathcal{N}_{node}^{r
}(i)=\mathcal{N}^1_{node}\{\mathcal{N}_{node}^{r
-1}(i)\}/[\{\cup_{q=1}^{r
-1}\mathcal{N}_{node}^{q}(i)\}\cup \{i \}].$$
\sloppy {For an edge $\{i,j\}$} the  set {$\mathcal{N}_{edge}^{1}(\{i,j\})$} of \textit{1-stage neighbouring edges} 
is the set of all edges which are incident to at least one of the nodes $i,j$.
Formally, 
$$\mathcal{N}_{edge}^{1}(\{i,j\})=\{\{k,l\}\in \delta^{+}(i)\cup \delta^{-}(i)\cup \delta^{+}(j)\cup \delta^{-}(j): \{k,l\}\neq \{i,j\} \}$$ with $\delta^{+}(\cdot),\delta^{-}(\cdot)$ denoting the sets of outgoing and incoming edges  of a node, respectively. 
For $r
\ge 2$ we similarly {denote} the set of $r
$-stage neighbouring edges of the edge $\{i,j\}$ 
as the set 
$$\mathcal{N}_{edge}^{r
}(\{i,j\})=\mathcal{N}^1_{edge}\{\mathcal{N}_{edge}^{r
-1}(\{i,j\})\}/[\{\cup_{q=1}^{r
-1}\mathcal{N}_{edge}^{q}(\{i,j\})\}\cup \{\{i,j\} \}].$$

{Let $L$ denote the maximum number of past time stamps,
and let $R_l$ denote} 
the maximum stage of neighbouring nodes \am{and neighbouring edges for each lag $l=1,\ldots,L$}. 
{
For $\boldsymbol{G_t}=(G_{1,t},\ldots,G_{K,t})$ \gr{denoting the time series on the $K$ nodes at time $t$,}
and $\boldsymbol{P_t}=(P_{1,t},\ldots,P_{M,t})$ \gr{denoting the time series on the $M$ edges at time $t$,} \am{assuming a labelling  $q : \mathcal{E}\rightarrow \{1,\ldots,M\}$ 
on the set of edges }
(with {an ordering} 
on the edge set), our new \gr{{\it GNAR-ex}}  model is}
\begin{eqnarray*}
    G_{i,t} &
    = &  \sum_{l=1}^{L}\left[ \alpha_l G_{i,t-l}+\sum_{r
    =1}^{
    {R_l} } \frac{\beta_{l,r
    }}{|\mathcal{N}_{node}^{r
}(i)|} \sum_{j\in N_{node}^{r
    }(i)}
    G_{j,t-l} \right. \\
     & &\left. 
     +\frac{\gamma_{l}}{|E_i|} \sum_{kp\in E_i}\left\{
     P_{kp,t-l}+\sum_{r
     =1}^{ {R_l} 
     }\frac{\delta_{l,r
     }}{|\mathcal{N}_{edge}^{r
}(\{k,p\})|} \sum_{mn\in N^{r
     }_{edge}(\{k,p\})}
     P_{mn,t-l}\right\}\right] +u_i^t
\label{eq1}\\
    P_{ij,t} &
    = & \sum_{l=1}^{L}\left[\alpha_l P_{ij,t-l}+\sum_{r
    =1}^{
     {R_l} }\frac{\beta_{l,r
     }}{| N_{edge}^{r
     }(\{i,j\})|} 
     \sum_{mn\in N_{edge}^{r
     }(\{i,j\})}
     P_{mn,t-l} \right. \\
    & &\left.  + \frac{\gamma_l}{2} {(G_{i,t-l}+G_{j,t-l})}+\sum_{r
    =1
    }^{{R_l} 
    }\frac{\delta_{l,r
    }}{|N^{r
    }_{node}(i)\cup N^{r
    }_{node}(j)/ [\{i\},\{j\}]| } \sum_{k\in 
    N^{r
    }_{node}(i)\cup N^{r
    }_{node}(j)/ [\{i\},\{j\}]} 
    G_{k,t-l}\right] +u_{ij}^t,
\label{eq2}
\end{eqnarray*}
with $E_i=\{\{i,j\}\cup\{j,i\}:j\in N_{node}^1(i) \}$ \kh{being} the set of edges which are incident to node $i$, $\alpha_l,\gamma_l$ being standard autoregressive parameters and $\beta_{l,r},\delta_{l,r}$ being network effect parameters, for $l=1,\ldots,L$ and $r=1,\ldots,R_l$.  
\kh{The error terms }$u_i^t$'s  and $u_{ij}^t$'s are \gr{assumed to be} i.i.d. centred normal with \gr{unknown} variance $\sigma^2$.
This setting gives a restricted {Gaussian} vector autoregressive (VAR) model for the multivariate time series comprising of \gr{a time series on the nodes and a time series on the edges of the static network.}
It inherits the theoretical guarantees for VAR models, such as asymptotic unbiasedness and asymptotic efficiency of least-squares estimates, \gr{see for example \cite{lutkepohl2005new}.}

{We compare the GNAR-x model to an ARIMA model, 
fitted individually to each \gr{node time series, ignoring the network structure.}
The ARIMA model has three components: an autoregressive component 
which indicates the size\gr{, $P$,} of lags on past values, the integrated component, \gr{with parameter $D$}  
which represents the \gr{order of} differencing of raw observations, and the moving average 
component, \gr{with parameter $Q$}, capturing the dependency on lagged residuals.
The general form of a {Gaussian} ARIMA($P,D,Q$) model is}
\begin{equation*}
    G_{i,t}^{'}=c+\sum_{p=1}^P\phi_{p}G_{i,t-p}^{'}+\sum_{q=1}^{Q}\theta_{q}\epsilon_{t-q}+\epsilon_t
\end{equation*}
where $G_{i,t}^{'}$ is the $D$-differenced time series \gr{for node} 
$i$, with $i=1,\ldots,K$, {and the $\epsilon_t$'s  are independent centred  normal with   variance $\sigma^2$}. For example, an ARIMA(0,1,0) model has the form
  $  G_{i,t}=c+G_{i,t-1}+\epsilon_t, $
{translating into assuming that the increments (the growth rates) are i.i.d.\,normal.} \gr{As a second comparison, we fit a vector autoregressive model using the Lasso-type VAR estimation process (HLAG) approach by \cite{nicholson2020high} \am{in synthetic data experiments. Due to the superior performance of ARIMA compared to VAR (HLAG) \gr{on the synthetic data}, in our real data application in Section \ref{sec:nowcastingGDP}   for comparisons to GNAR-ex we only consider the ARIMA model}.}

\am{The unrestricted VAR model 
for the multivariate time series data set comprising of $M+K$ time series}
\begin{eqnarray*}
Y_{n,t}&
    = 
    \begin{cases}
      P_{n,t}, & \text{if}\ n=1,\ldots,M \\
      G_{n,t}, & \text{if}\ n=M+1,\ldots,M+K
    \end{cases}
\end{eqnarray*}
has the form 
 \begin{eqnarray*}
    Y_{n,t} & =& v_n+\alpha_{n_{1,1}}Y_{1,t-1}+
    \ldots+\alpha_{n_{M+K,1}}Y_{M+K,t-1}+\ldots+\alpha_{n_{1,L}}Y_{1,t-L}+\ldots+\alpha_{n_{M+K,L}}Y_{M+K,t-L}  
    +u_n^t
. 
\end{eqnarray*}

\section{Simulation results for GNAR-ex}
\label{sec:simulations}
To validate the model and  
test its prediction performance, we simulate nodal and edge time series 
\kh{for networks with different topologies} and implement the GNAR-ex model on the simulated time series.
As in \citet{mantziou2023gnar}, we consider three \gr{commonly used random} network \gr{models,} 
namely an Erd\H{o}s-R\'enyi (ER), a Stochastic Block Model (SBM) and a geometric random graph, specifically a Random Dot Product graph (RDP).
The directed networks considered in our experiments have 20 nodes with \gr{parameters set to achieve} a density of approximately 0.4. 
We simulate nodal and edge time series using the GNAR-ex model under the parameter specifications presented in Table \ref{simreglag1}. \am{To describe each model considered in our experiments, \gr{the} notation GNAR-ex($L$,[$R_1,\ldots,R_L$]) denotes the GNAR-ex model with  maximum lag $L$ and maximum stages of neighbours $R_l$,  for  lags $l=1,\ldots,L$. }We fit the GNAR-ex model on the simulated time series to make inferences for the model parameters. To evaluate the estimation performance of the model, we obtain the RMSE of the estimated coefficients with respect to their true values
, and calculate the proportion of times that the true values lie within the 95\% confidence interval. For each simulation regime, we repeat the experiment 50 times for 50 different seeded simulated data sets. In Tables \ref{simreg1reslag1}, \ref{simreg2reslag1} and \ref{simreg3reslag1} we present the results for each simulation regime. 

We observe that under all three simulation regimes and different network structures, we obtain accurate estimates of the true size of the model parameters, as indicated by the small RMSE values and high coverage close to nominal level of 95\%. \am{In particular, 
\gr{the} estimation performance for the autoregressive parameters $\boldsymbol{\alpha}$ and $\boldsymbol{\gamma}$ \gr{is better than that} for the network autoregressive parameters $\boldsymbol{\beta}$ and $\boldsymbol{\delta}$, as indicated by the smaller RMSE compared to the RMSE, across all simulation regimes. This result suggests that making inferences for the network parameters is more challenging; nonetheless, we obtain a relatively small RMSE with respect to the true size of both the autoregressive and network effect parameters. 
The estimation performance is 
similar \mc{across the different network models  considered; however, we observe} that for the ER networks in the simulation regime with the highest complexity (simulation regime 3), the coverage of 95\% confidence intervals is slightly smaller for the autoregressive parameters at lag 2 \gr{than for the network effect parameters}.}

\begin{table}[htb!]
\centering
\scalebox{0.8}{
\begin{tabular}{llcccc}
\kh{\textbf{Regime}}&                                          & $\boldsymbol{\alpha}$ &  $\boldsymbol{\beta}$               & $\boldsymbol{\gamma}$       & $\boldsymbol{\delta}$                              \\ \hline
\textbf{1} & \textbf{GNAR-ex(1,{[}1{]})}     & 0.2        & 0.2                       & 0.3        & 0.2                          \\ 
\textbf{2} & \textbf{GNAR-ex(1,{[}2{]})}     & 0.2        & (-0.2,0.1)                & 0.1       & (0.05,-0.2)               \\ 
\textbf{3} & \textbf{GNAR-ex(2,{[}2,2{]})} & (-0.1,0.3) & ((0.1,-0.2),(-0.02,0.03)) & (0.01,0.01) & ((0.02,-0.01),(-0.02,0.01)) \\ 
\end{tabular}
}
\caption{Simulation regimes for 20-node networks.}\label{simreglag1}
\end{table}

\begin{table}[htb!]
\centering
\scalebox{0.8}{
\begin{tabular}{lcccccc}
                     & \multicolumn{1}{l}{\textbf{Regime 1}} & \multicolumn{1}{l}{\textbf{}} & \multicolumn{1}{l}{\textbf{}}    & \multicolumn{1}{l}{\textbf{}} & \multicolumn{1}{l}{\textbf{}}    & \multicolumn{1}{l}{}              \\
                     & \multicolumn{1}{l}{\textbf{ER}}       & \multicolumn{1}{l}{\textbf{}} & \multicolumn{1}{l}{\textbf{SBM}} & \multicolumn{1}{l}{\textbf{}} & \multicolumn{1}{l}{\textbf{RDP}} & \multicolumn{1}{l}{}              \\
                     & \multicolumn{1}{l}{\textbf{Coverage}} & \textbf{RMSE}                 & \textbf{Coverage}                & \textbf{RMSE}                 & \textbf{Coverage}                & \multicolumn{1}{l}{\textbf{RMSE}} \\ \hline

\textbf{$\boldsymbol{\alpha_1}$}    & 0.96                                     & 0.005                         & 0.96                             & 0.005                         & 0.92                             & 0.005                             \\
\textbf{$\boldsymbol{\beta_{1,1}}$}  & 0.98                                   & 0.02                          & 0.9                             & 0.02                          & 0.96                             & 0.02                              \\
\textbf{$\boldsymbol{\gamma_1}$}    & 0.96                                  & 0.007                         & 0.94                             & 0.008                         & 0.92                              & 0.008                             \\
\textbf{$\boldsymbol{\delta_{1,1}}$} & 0.94                                  & 0.02                          & 0.98                                & 0.01                          & 0.96                             & 0.02                             
\end{tabular}}
\caption{RMSE of estimated coefficients and coverage of 95\% confidence intervals for simulated network time series under simulation regime 1, for different network structures (ER, SBM, RDP).}\label{simreg1reslag1}
\end{table}

\begin{table}[htb!]
\centering
\scalebox{.8}{
\begin{tabular}{lcccccc}
                     & \multicolumn{1}{l}{\textbf{Regime 2}} & \multicolumn{1}{l}{\textbf{}} & \multicolumn{1}{l}{\textbf{}}    & \multicolumn{1}{l}{\textbf{}} & \multicolumn{1}{l}{\textbf{}}    & \multicolumn{1}{l}{}              \\
                     & \multicolumn{1}{l}{\textbf{ER}}       & \multicolumn{1}{l}{\textbf{}} & \multicolumn{1}{l}{\textbf{SBM}} & \multicolumn{1}{l}{\textbf{}} & \multicolumn{1}{l}{\textbf{RDP}} & \multicolumn{1}{l}{}              \\
                     & \multicolumn{1}{l}{\textbf{Coverage}} & \textbf{RMSE}                 & \textbf{Coverage}                & \textbf{RMSE}                 & \textbf{Coverage}                & \multicolumn{1}{l}{\textbf{RMSE}} \\ \hline

\textbf{$\boldsymbol{\alpha_1}$}    & 0.96                                  & 0.005                         & 0.96                             & 0.004                         & 0.96                             & 0.005                             \\
\textbf{$\boldsymbol{\beta_{1,1}}$}  & 0.92                                  & 0.02                          & 0.9                             & 0.03                          & 0.9                            & 0.03                              \\
\textbf{$\boldsymbol{\beta_{1,2}}$}  & 0.94                                  & 0.03                          & 0.94                             & 0.03                          & 0.96                             & 0.04                              \\
\textbf{$\boldsymbol{\gamma_1}$}    & 0.96                                  & 0.007                         & 0.94                             & 0.008                         & 0.98                             & 0.008                             \\
\textbf{$\boldsymbol{\delta_{1,1}}$} & 0.96                                  & 0.03                          & 0.98                             & 0.02                          & 0.94                             & 0.03                              \\
\textbf{$\boldsymbol{\delta_{1,2}}$} & 0.94                                  & 0.02                          & 0.92                             & 0.02                          & 0.9                              & 0.03                          
\end{tabular}}
\caption{RMSE of estimated coefficients and coverage of 95\% confidence intervals for simulated network time series under simulation regime 2, for different network structures (ER, SBM, RDP).}\label{simreg2reslag1}
\end{table}

\begin{table}[htb!]
\centering
\scalebox{0.8}{
\begin{tabular}{lcccccc}
                     & \multicolumn{1}{l}{\textbf{Regime 3}} & \multicolumn{1}{l}{\textbf{}} & \multicolumn{1}{l}{\textbf{}}    & \multicolumn{1}{l}{\textbf{}} & \multicolumn{1}{l}{\textbf{}}    & \multicolumn{1}{l}{}              \\
                     & \multicolumn{1}{l}{\textbf{ER}}       & \multicolumn{1}{l}{\textbf{}} & \multicolumn{1}{l}{\textbf{SBM}} & \multicolumn{1}{l}{\textbf{}} & \multicolumn{1}{l}{\textbf{RDP}} & \multicolumn{1}{l}{}              \\
                     & \multicolumn{1}{l}{\textbf{Coverage}} & \textbf{RMSE}                 & \textbf{Coverage}                & \textbf{RMSE}                 & \textbf{Coverage}                & \multicolumn{1}{l}{\textbf{RMSE}} \\ \hline

\textbf{$\boldsymbol{\alpha_1}$}    & 0.98                                  & 0.005                         & 0.96                             & 0.005                         & 0.96                             & 0.005                             \\
\textbf{$\boldsymbol{\beta_{1,1}}$}  & 0.96                                  & 0.03                          & 0.96                             & 0.02                          & 0.94                             & 0.02                              \\
\textbf{$\boldsymbol{\beta_{1,2}}$}  & 0.94                                  & 0.03                          & 0.94                             & 0.03                          & 0.96                             & 0.03                              \\
\textbf{$\boldsymbol{\gamma_1}$}    & 0.98                                  & 0.006                         & 0.94                             & 0.007                         & 0.96                             & 0.008                             \\
\textbf{$\boldsymbol{\delta_{1,1}}$} & 0.94                                  & 0.03                          & 0.96                             & 0.02                          & 0.92                             & 0.02                              \\
\textbf{$\boldsymbol{\delta_{1,2}}$} & 0.94                                  & 0.02                          & 0.94                             & 0.02                          & 0.96                             & 0.02                              \\
\textbf{$\boldsymbol{\alpha_2}$}    & 0.88                                  & 0.006                         & 0.96                             & 0.004                         & 1                                & 0.004                             \\
\textbf{$\boldsymbol{\beta_{2,1}}$}  & 0.98                                  & 0.03                          & 0.94                             & 0.02                          & 0.96                             & 0.02                              \\
\textbf{$\boldsymbol{\beta_{2,2}}$}  & 0.96                                  & 0.03                          & 0.94                             & 0.03                          & 0.92                             & 0.03                              \\
\textbf{$\boldsymbol{\gamma_2}$}    & 0.86                                  & 0.009                         & 0.96                             & 0.008                         & 0.94                             & 0.009                             \\
\textbf{$\boldsymbol{\delta_{2,1}}$} & 0.98                                  & 0.02                          & 1                             & 0.02                          & 0.94                             & 0.03                              \\
\textbf{$\boldsymbol{\delta_{2,2}}$} & 0.98                                  & 0.02                          & 0.92                             & 0.02                          & 0.98                             & 0.02                             
\end{tabular}}
\caption{RMSE of estimated coefficients and coverage of 95\% confidence intervals for simulated network time series under simulation regime 3, for different network structures (ER, SBM, RDP).}\label{simreg3reslag1}
\end{table}
\newpage

We further explore the predictive performance of the GNAR-ex model and compare it to two baseline models, namely an ARIMA model fitted individually to each time series using the \texttt{auto.arima} function that allows model selection with BIC/AIC criteria, and a vector autoregressive model using \gr{the} Lasso-type VAR estimation process (HLAG) approach by \cite{nicholson2020high}. We consider the third simulation regime from Table 1, which is the most complex regime considered in our simulation experiments. 
\gr{Again} we consider 20-node networks with different structures (SBM,ER,RDP) and density approximately 0.4. 

\begin{figure}[ht!]
    \centering
    \includegraphics[scale=.45]{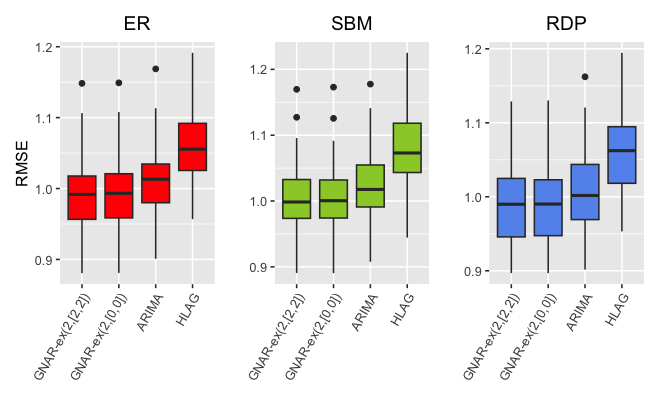}
    \caption{Prediction performance of GNAR-ex model with and without neighbourhood structure, ARIMA model and VAR model using the HLAG approach by \cite{nicholson2020high}, for all time series (nodal and edge). 
    }
    \label{edgenodeRMSEpred}
\end{figure}

We explore the predictive performance of the different models by fitting each model on the simulated network time series under simulation regime 3, excluding the last time stamp, and predict the last time stamp using the fitted model. To evaluate the prediction accuracy, we use the RMSE of the predicted value with respect to the true value of the last time stamp. We repeat this experiment 50 time\mc{s} for 50 different seeded data sets, and present the distribution of the RMSE across repetitions. The RMSE of the predictions combined for both the nodal and edge time series are presented in Figure \ref{edgenodeRMSEpred}. As the main task for the real data application is forecasting of GDP values (nodal time series), in Figure \ref{nodeRMSEpred} in Appendix \ref{sec:appendix} we present the RMSE of the prediction only for the nodal time series.

The results suggest that the GNAR-ex model 
\gr{performs better than} the ARIMA and VAR model across the different network structures.
\am{
We see that the GNAR-ex model with and without neighbour structure have similar prediction performance. This 
\gr{may be related to} the maximum size of neighbourhoods considered (2 in this experiment) and  the size of networks;
 in \cite{mantziou2023gnar} it is found that 
 the neighbour effect 
 plays a more vital role in settings with higher stages of neighbours.} 
\am{Due to \mc{the} VAR model (HLAG) having the less competitive performance to the GNAR-ex, for the rest of the simulaiton experiments in this section and in the real data section \ref{sec:nowcastingGDP}, for comparisons we consider only the ARIMA model \gr{family}.}

\am{It is important to note that \gr{the GNAR-ex model can be fitted in reasonable time;} 
the run time for the GNAR-ex model for 50 repetitions under the most complex regime, simulation regime 3, is 6 minutes, run in RStudio on MacBook Pro M3 machine.}

\subsection{Model averaging}

In this section, we investigate the predictive performance of the model in a setting that involves uncertainty with respect to model choice, which is typical when dealing with real data applications. 
Model uncertainty in our case arises from the specification of the stages of neighbours and lag sizes for the GNAR-ex model when we do not know the true data generative mechanism. One way to 
\gr{address} this type of uncertainty arising in real data applications, is to fit various models and obtain an average forecast across the different models \citep{fletcher2018model}. 

To test the predictive performance after 
averaging different GNAR-ex models, we consider the following experiment.
First, we consider the simulation regime 3 of Table \ref{simreglag1}, which involves nodal and edge time series generated under the assumption of a 2-stage neighbour structure. 
For this regime, we simulate 50 different seeded data sets, 
estimate the parameters of the GNAR-ex model with 1-stage neighbours and different lag specifications excluding the last time stamp, and predict the last time stamp using the average forecast across the GNAR-ex models specified. 
We evaluate the forecast resulting from model averaging (MA) 
using the RMSE and compare the results with the ARIMA model similarly to previous section.
We note here that, with this experiment, we also test the performance of GNAR-ex under model mis-specification, since the true data generating mechanism is a 2-stage neighbour model, while we use 1-stage neighbour models to perform MA.

\begin{figure}[ht!]
    \centering
    \includegraphics[scale=.4]{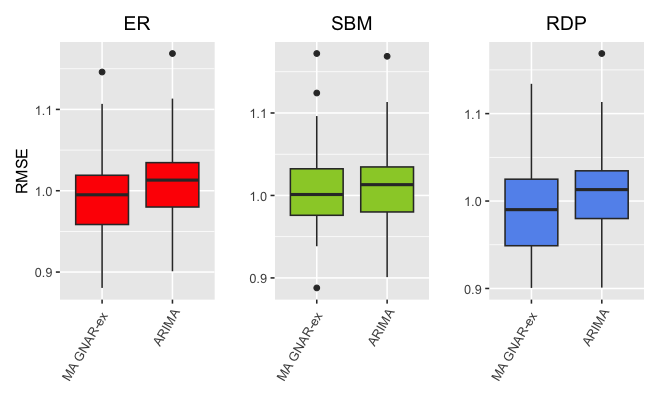}
    \caption{Prediction performance for all time series (nodal and edge) from MA across various lag sizes ($l=1,\ldots,9$) the GNAR-ex model with 1-stage neighbours compared to best fitting ARIMA model, for simulation regime 3 in Table \ref{simreglag1}.}
    \label{edgenodeRMSEpredMA}
\end{figure}
\begin{figure}
\centering
\vfill
    \includegraphics[scale=.4]{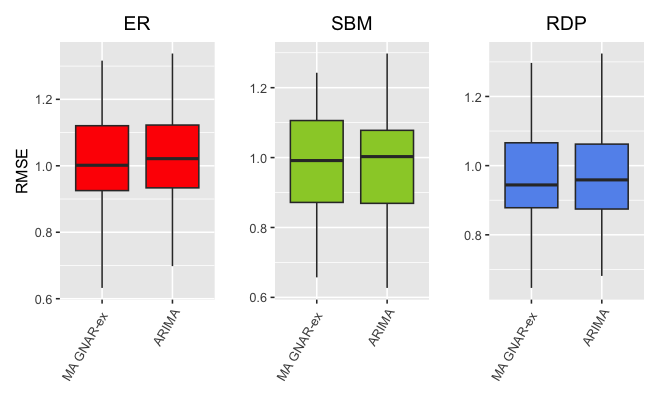}
    \caption{Prediction performance for nodal time series from MA across various lag sizes ($l=1,\ldots,9$) the GNAR-ex model with 1-stage neighbours compared to best fitting ARIMA model, for simulation regime 3 in Table \ref{simreglag1}. }
    \label{nodeRMSEpredMA}
\end{figure}

Figure \ref{edgenodeRMSEpredMA} shows the distribution of the RMSE of the MA forecast across repetitions, for the edge and nodal time series combined. We observe  that despite the model mis-specification and MA performed, \mc{the GNAR-ex model is still}  outperforming the best fitting ARIMA model. Similarly, we present in Figure \ref{nodeRMSEpredMA} the results considering only the nodal time series. 
\gr{Keeping in mind that only 20 of the around 180 time series in the models are nodal time series, the prediction performance of the nodal time series, shown in Figure \ref{nodeRMSEpredMA} is not as granular as that for the whole set of time series shown in Figure \ref{edgenodeRMSEpredMA}, but s}till the MA GNAR-ex model is performing well compared to the best fitting ARIMA model, with the median RMSE of the MA GNAR-ex being slightly smaller than the median RMSE of the best fitting ARIMA model. 

Since the GNAR-ex model is a restricted VAR model with Gaussian noise assumption, we are able to obtain forecast intervals,  which are particularly useful in real data applications to account for the uncertainty in the forecast \citep{lutkepohl2005new}. We test the accuracy of the forecast intervals from the MA GNAR-ex model by obtaining the union of forecast intervals across the GNAR-ex models with 1-stage neighbours and different lag sizes.
Specifically, we calculate the proportion of nodes for which the true value of the last time stamp lies within the union forecast interval, across the 50 repetitions. 
We observe that across all network structures considered, at least 85\% of the last time stamps of the nodal time series lie within the union forecast interval, indicating that \mc{the} union forecast interval from MA is a reliable summary to use for the real data application. 

In Table \ref{forecastunion}, we show the proportion of repetitions (out of the 50) for which  85\% (17 out of 20), 90\% (18 out of 20), 95\% (19 out of 20) and 100\% (20 out of 20) of the nodal time series lie within the union forecast interval. 
\gr{To gauge these results, we}
calculate the probability of observing at least 19 nodes out of the 20 (95\% of nodes) lying within the union forecast interval under the assumption that the number of nodes lying in 95\% forecast interval follows a Binomial distribution with probability of success 0.95. By simple calculations we obtain that the probability of at least 95\% of the nodes to lie within the union forecast interval is 0.74. 

Comparing to the results from the repetitions of our experiment, we observe that for the ER model, at least 19 nodes lie within the union forecast interval for 76\% of repetitions, while for SBM and RDP is 88\% and 90\% of repetitions respectively; thus we see that for all cases 
\gr{the performance is}
higher than the theoretical probability of 0.74. In a similar manner, the theoretical probability of at least 85\%, 90\% and 100\% of the nodes to lie within the union forecast interval is 0.98, 0.92 and 0.38, respectively, indicating that in all cases the performance is higher than the theoretical probability. We note here that this is not an exact calculation, due to the independence assumption and the conservative union of forecast intervals obtained, but rather serves as a point of reference for comparing the results across repetitions to understand the model performance.

\begin{table}[htb!]
\centering
\begin{tabular}{lllll}
             & \multicolumn{4}{l}{\textbf{Forecast inclusion}} \\
             & 17       & 18        & 19     & 20        \\ \hline
\textbf{ER}  & 0.02       & 0.22       & 0.34      & 0.42      \\
\textbf{SBM} & 0.02       & 0.1        & 0.42      & 0.46      \\
\textbf{RDP} & 0.04       & 0.06       & 0.24      & 0.66  
\end{tabular}
\caption{Proportion of repetitions for which  
\gr{exactly 17 (85\%), 18 (90\%),  19 (95\%) or 20 (100\%) of the 20} 
nodal time series (true last time stamp) lie within the union 95\% forecast interval, for different network structures (ER,SBM,RDP).}\label{forecastunion}
\end{table}

We repeat the above experiment of MA for various lags of a 1-stage GNAR-ex model, for data simulated under regime 1 in Table \ref{simreglag1} assuming 1-stage neighbour structure, to explore model performance in a simpler setting where there is no model mis-specification. As expected, in this simpler setting we have an improved forecasting performance of the MA GNAR-ex compared to the previous setting. The results are presented in Appendix \ref{sec:appendix}.

\section{Nowcasting UK GDP}

In this section, we describe how 
we obtain a  
network from the data as well as the  
{nowcasting} task.
We propose \mc{a} sparsification of the network because 
high correlations between GDP and payments time series can 
affect the precision of the parameter estimates. 
\label{sec:nowcastingGDP}

\subsection{Network sparsification}
In the first step, we remove industries (nodes) from the network with respect to (1) knowledge about issues with the payment data and (2) economic intuition, namely no expected relevance for GDP fluctuations. The industries removed from the network at this first step are the following: 
\begin{itemize}
    \item 
O84 Public administration, \item G45 Wholesale vehicles, \item G46 Wholesale except vehicles,\item  G47 Retail except vehicles, \item Q86 Human health, \item Q87-88 Residential care.
\end{itemize}

The reasoning is as follows.
\gr{Early exploration of the payment data set indicated that}
\kh{
the Public administration sector suffers from inconsistencies between payment \gr{data} and GDP accounting \gr{data}  with regard to recording of economic transactions.} 
Further, the classification \gr{in the payment data} is conceptually different from the practices \gr{often} used in official statistics \citep{hotte2024}. 
The retail \gr{and} wholesale industries are intermediary sectors that ``destroy'' the direct link between the buying and selling industry (however, they should be 2-stage neighbours).
Health- and care-related spendings are excluded for conceptual reasons, as they are expected to be unrelated to GDP fluctuations. 

{The second step for sparsifying the network involves removal of payments (edges) that are highly correlated to GDP time series. Specifically, we remove edges corresponding to payments time series that have cross-industry Pearson's correlation to GDP nodal time series above 0.4 or below -0.4, thus keeping \gr{only} time series with weak correlation  \gr{in the sense of} \citep{evans1996straightforward}. 
The resulting network has 32 nodes and 231 edges.}

\subsection{Approach}

To examine the {nowcasting}  
accuracy of our model across the different data releases, 
we use each \kh{of the 9 }GDP data releases as a training sample to fit our model, nowcast the next (monthly) time stamp, 
and evaluate the nowcasting accuracy with data in subsequent releases. Each GDP data release  provides industry GDP data with a lag of 6 months. For example, the data release of December 2021 includes industry GDP values up to June 2021. In our setup, we use this data as training sample to fit our model and {nowcast} 
the next time stamp, 
July 2021. To evaluate the {nowcasting}
accuracy, we compare our  
{nowcast} to the July 2021 industry GDP values released in the subsequent data release of March 2022. We repeat the same procedure for the 9 \gr{ONS} data releases \gr{since \am{December} 2021}.

{For the time series of growth rates, we apply a seasonal-trend decomposition (STL) using the R function \texttt{stl} and fit the GNAR-ex model on the residuals.
We evaluate the forecasting accuracy on the original scale of the GDP value by obtaining the total GDP (sum of industry-level GDP values) and using the relative error ${|\text{forecast-actual}|}/ {\text{actual}}$ as a measure of {nowcasting}  
performance.}
\am{{For comparison}, we fit an ARIMA model on \gr{the} raw GDP time series  \gr{for each industry} and allow the parameter selection to be performed through the AIC criterion using the R function \texttt{auto.arima}. \gr{We also fit an ARIMA(0,1,0) model \gr{as this was identified as} \am{
the best fitting model when using \texttt{auto.arima}}.}
}

\subsection{Results}

In this section, we present the results after the implementation of the GNAR-ex model on the pre-processed data, across different GDP releases. We further perform model averaging 
and compare to an ARIMA model. Summaries of industry-level results after model averaging are also presented in this section.
We evaluate the model performance after rescaling predictions to the original scale of the raw time series.

\subsubsection{Results across data releases}

\begin{figure}[htb!]
    \centering
    \includegraphics[scale=.35]{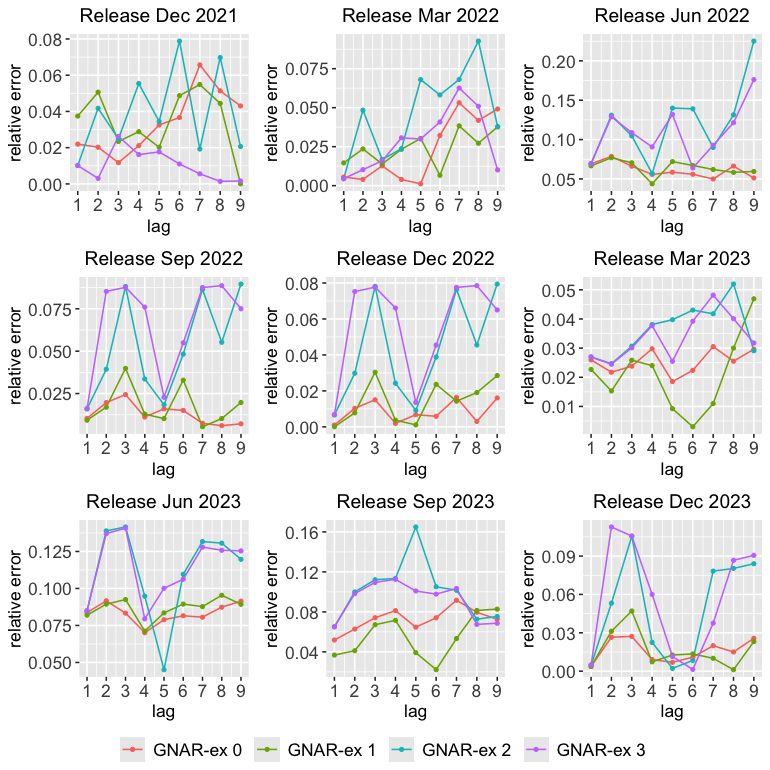}
    \caption{Relative Error of total GDP for GNAR-\kh{e}x-$R_l$ model with lag $L=1,\ldots,9$ and neighbour stages $R_l=0,\ldots,3$, across releases.} 
    \label{relerr_releases_growth_nodedgerem}
\end{figure}

In Figure \ref{relerr_releases_growth_nodedgerem} we report the relative error of total GDP forecast using the GNAR-ex model for a range of lags and stages of neighbours. We observe high variability of the relative error for different lags and stages of neighbours. However, the GNAR-ex model with 1-stage neighbours is the best performing model across most releases. We also notice close performance of the GNAR-ex model with 0-stage neighbours. Even though the 0-stage neighbours implies the absence of network effect parameters in the model, still the 0-stage GNAR-ex model includes \gr{network information through the}  
\kh{in- and outgoing payment flows observed on the directly incident edges.}

In Table \ref{bestmod_gnar_ar_arima_nodedgerem}, we summarise the results across releases, by obtaining the best performing GNAR-ex model for each release and comparing to an ARIMA model fitted individually to each industry-level GDP time series, allowing for lag, differencing and moving average specifications to be determined with information criteria (AIC), using the \texttt{auto.arima} function in R.
The results from fitting an \texttt{auto.arima} on the GDP time series indicate  that an ARIMA(0,1,0) is the best fitting model, thus we further compare to an ARIMA(0,1,0) fitted individually to all time series in last column of Table \ref{bestmod_gnar_ar_arima_nodedgerem}. 
We observe that the GNAR-ex model allows for better forecasts of total GDP across all releases when compared to the best fitting ARIMA model.
{Compared to an ARIMA(0,1,0), a  GNAR-ex model still performs better for the majority of data releases}, with two cases resulting in similar error.

Here the ARIMA(0,1,0) model often has smaller relative error than the ``best-fitting'' ARIMA model. This discrepancy arises because the best-fitting ARIMA model is selected based on AIC, and not on the prediction error. 
It may be useful to keep in mind that the GDP \kh{data used in this work is itself an early estimate, based on short-term available survey information and autoregressive elements.}

\begin{table}[htb!]
\centering
\begin{tabular}{lrrrrr}
  \hline Release
 & GNAR-ex lag & GNAR-ex stage & GNAR-ex error & auto.ARIMA error & ARIMA(0,1,0) error \\ 
  \hline
 Dec 2021 & 9 & 1 & \textbf{0.00004} & 0.01  & 0.007 \\ 
   Mar 2022 & 5 & 0 & \textbf{0.001} & 0.08  & \textbf{0.001} \\ 
 Jun 2022 & 4 & 1 & \textbf{0.04} & 0.05  & \textbf{0.04 } \\ 
 Sep 2022 & 7 & 1 & \textbf{0.005} & 0.03  & 0.01 \\ 
 Dec 2022 & 1 & 1 & \textbf{0.00004}  & 0.08  & 0.004  \\ 
 Mar 2023 & 6 & 1 & \textbf{0.003} & 0.03  & 0.01 \\ 
 Jun 2023 & 5 & 2 & 0.04 & 0.09  & \textbf{0.003} \\ 
 Sep 2023 & 6 & 1 & 0.02 & 0.09  & \textbf{0.007} \\ 
 Dec 2023 & 8 & 1 & \textbf{0.001} & 0.10  & 0.007 \\ 
   \hline
\end{tabular}
\caption{Best performing GNAR-\gr{e}x model with respect to forecasting accuracy, best performing ARIMA with respect to information AIC criterion and ARIMA (0,1,0) model. Columns GNAR-ex lag and GNAR-ex stage indicate the lag and stages of neighbours for which \mc{the}  relative error is minimised across releases. \mc{The models} with minimum error across releases are in bold.}
\label{bestmod_gnar_ar_arima_nodedgerem}
\end{table}

As the GNAR-ex model with edge and node time series is a restricted VAR model, it allows theoretical guarantees to obtain forecast intervals. In Table \ref{predint_dec21}, we report the proportion of times that the true industry-level GDP values (node-level) lie within the 95\% forecast intervals obtained from the GNAR-ex model, for the December 2021 release \kh{chosen as an example}, 
and for the different lag and stages. In Appendix \ref{sec:app_realdata} we  
report the forecast interval results for the other the releases. We  
\gr{stress} that we do not 
expect the proportions obtained to be close to nominal level of 95\% of the forecast intervals, as the proportions result from the number of industries (nodes) \gr{with} GDP values lying within the forecast intervals, rather than from multiple repetitions of our model on different samples. We see an overall good performance of the GNAR-ex model on forecasting industry-level GDP values as indicated by the close to 1 proportion of industry-level GDP values lying within the 95\% forecast intervals. However, we also observe variability of the performance across releases and across lags and stages of neighbours. Overall, \gr{better} 
performance is observed for smaller lag specifications.

\begin{table}[ht]
\centering
\begin{tabular}{rrrrr}
  \hline
lag & stage 0 & stage 1 & stage 2 & stage 3 \\ 
  \hline
 1 & 0.97 & 0.97 & 0.97 & 0.97 \\ 
  2 & 0.97 & 0.97 & 0.91 & 0.97 \\ 
  3 & 0.94 & 0.91 & 0.94 & 0.94 \\ 
  4 & 0.94 & 0.88 & 0.94 & 0.94 \\ 
  5 & 0.94 & 0.97 & 0.97 & 0.91 \\ 
  6 & 0.91 & 0.94 & 0.94 & 0.91 \\ 
  7 & 0.84 & 0.91 & 0.91 & 0.91 \\ 
  8 & 0.91 & 0.91 & 0.84 & 0.91 \\ 
  9 & 0.91 & 0.91 & 0.94 & 0.88 \\ 
   \hline
\end{tabular}
\caption{
\gr{Proportion of industries for which the}  true industry level GDP values (original scale) lie within the 95\% forecast intervals, 
for each lag and stages of neighbours, for the  December 2021 data release.}
\label{predint_dec21}
\end{table}

\subsubsection{Model averaging}

Results from the GNAR-ex model can vary across different specifications of lag size and stages of neighbours, and model choice according to performance can be complex. To deal with model uncertainty, particularly arising when dealing with real data, we perform \gr{model averaginng} (\kh{MA}).  
The results obtained from fitting the GNAR-ex model among the different releases suggest that the 1-stage neighbour GNAR-ex model \gr{tends to perform better than}
other stage specifications across releases. 
Due to this observation and \am{results obtained from synthetic data experiments in previous section,} we 
average the GNAR-ex model with 1-stage neighbours across the different lags $L=1,\ldots,9$. 
\gr{Specifically,} we take the average of the total 
GDP forecast across lags $\frac19 {\sum_{i=1}^9 \text{GDPtotal}_{i,1}}$ where we equally weight 
the total GDP forecasts across the models, for each release, and calculate the relative error.
In Table \ref{ma1_autoarima}, we report the relative error after \kh{MA} 
and compare to the best fitting ARIMA using AIC criterion and the ARIMA(0,1,0).

\begin{table}[htb!]
\centering
\begin{tabular}{lrrr}
  \hline Release
 & MA all GNAR-ex stage 1 &  auto.ARIMA & ARIMA(0,1,0) \\ 
  \hline Dec 2021 & \textbf{0.00062}& 0.01 & 0.007 \\ 
   Mar 2022 &0.0073 & 0.08 & \textbf{0.001}\\ 
 Jun 2022 & 0.064 & 0.05 & \textbf{0.04}\\ 
 Sep 2022 & \textbf{0.0096} & 0.03 & 0.01 \\ 
 Dec 2022 & \textbf{0.00053} & 0.08 &0.004\\ 
 Mar 2023 & 0.02 & 0.03 &\textbf{0.01}\\ 
 Jun 2023 & 0.087 & 0.09 &\textbf{0.003}\\ 
 Sep 2023 & 0.055 & 0.09&\textbf{0.007} \\ 
 Dec 2023 & \textbf{0.0032} & 0.1 &0.007\\ 
   \hline
\end{tabular}
\caption{Model averaging over different lags, for GNAR-ex with 1-stage neighbours, and comparison to the 
ARIMA with 
\gr{lowest} AIC. 
The model
with minimum error across releases is in bold.}
\label{ma1_autoarima}
\end{table}

Despite the anticipated increase in the relative error due to MA, 
the MA GNAR-ex model outperforms the best fitting ARIMA model for most releases (8 out of 9) and the ARIMA(0,1,0) model for 4 out of the 9 releases. In Table \ref{ma1_unionforecast} we also report the proportion of industries (nodes) for which their true values lie within the union forecast intervals obtained after MA. For most releases, we observe that  high proportion of industries  for which true values lie within the forecast intervals (for 6 out of 9 releases 100\% of nodes), with the lowest value being for the data release of September 2023 which is also a release for which an ARIMA(0,1,0) has better forecasting performance. 
These results combined with 
the insights from the  synthetic data experiments suggest that applying MA for the GNAR-ex model is a competent way to retain good forecasting performance while  
\gr{addressing} model uncertainty.

\begin{table}[htb!]
\centering
\begin{tabular}{lr}
  \hline Release
 & Forecast inclusion\\ 
  \hline
 Dec 2021  & 0.97\\ 
 Mar 2022 & 1 \\ 
 Jun 2022 & 0.81 \\ 
 Sep 2022 & 1 \\ 
 Dec 2022 & 1\\ 
 Mar 2023 & 1 \\ 
 Jun 2023 & 1 \\ 
 Sep 2023 & 0.78 \\ 
 Dec 2023 & 1\\ 
   \hline
\end{tabular}
\caption{Proportion of nodes (industries) for which the true value lies within the union forecast interval from MA over different lags, for GNAR-ex with 1-stage neighbours, across releases.}
\label{ma1_unionforecast}
\end{table}

An advantage of the GNAR-ex model is that it \mc{allows} inferences on the industry level with respect to  
\kh{supply chain dynamics embodied in the network time series of payments. Such granular information cannot be inferred from aggregate approaches to GDP nowcasting}. In the next section, we take a closer look at the GNAR-ex forecasting performance at industry level, across releases, \gr{to illustrate how the GNAR-ex model can yield a}
better understanding  \gr{of }forecasting challenges on a more granular level and \gr{can help gain}  
economic insights.

\subsection{Industry-level results after model averaging}

\begin{figure}[htb!]
    \centering
    \includegraphics[scale=.4]{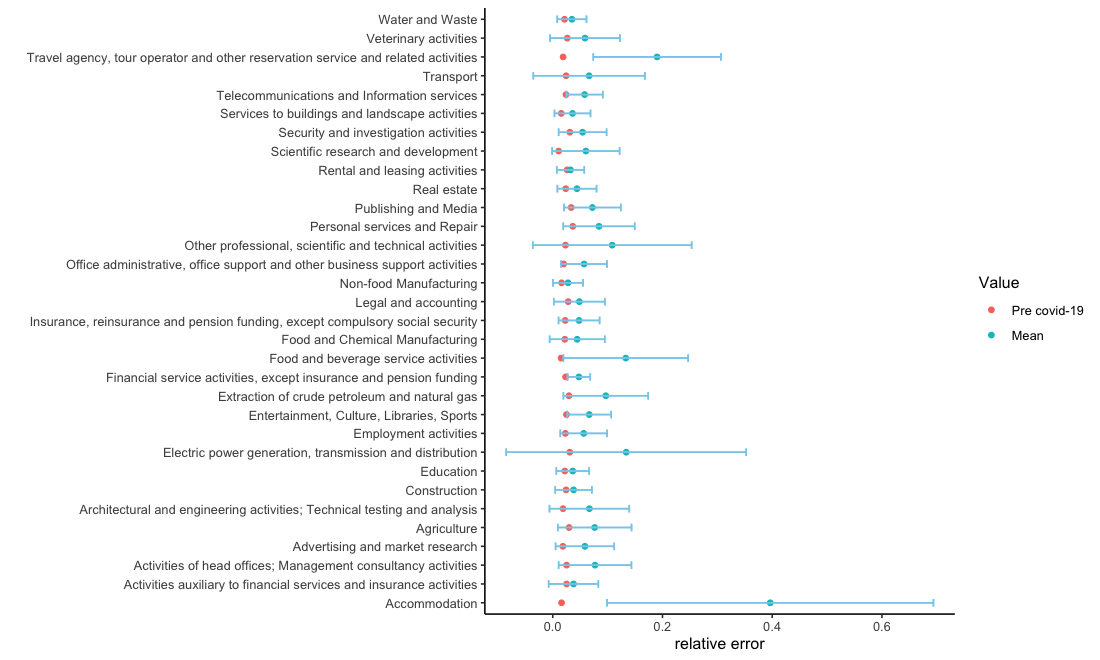}
    \caption{Industry-level mean and sd (blue) of relative error across 9 data releases, after model averaging across lags $=1,\ldots,9$, for stage-1 neighbours and industry-level relative error (red) for pre Covid-19 data after model averaging across lags $=1,\ldots,9$, for stage-1 neighbours.}
    \label{indrelerrcov}
\end{figure}

To better understand the forecasting accuracy of the MA GNAR-ex model per industry, we visualise the mean relative error per industry across the 9 releases, along with the standard deviation (sd) as indicated by the error bars in Figure \ref{indrelerrcov}. Some industries, such as ``Accommodation'' and ``Travel Agency'', 
\gr{have} high mean and sd values of the relative error across releases,  suggesting that the forecasting task can be particularly challenging for specific industries. \gr{This could} 
subsequently affect the forecasting accuracy of the total GDP, seen in the previous section. 

Specific industries such as ``Accommodation'' and ``Travel Agency''
\kh{were among the most}
affected industries by \kh{containment policies during} the Covid-19 pandemic \citep{nicola2020socio}
,  a period which is covered in the multivariate time series of payments and GDP.  
Motivated by this observation, we perform the same MA procedure \gr{as above, but now} using as training sample \gr{only} the latest release (Dec 2023) of the multivariate time series up to and including December 2019 \gr{as these data would be the most accurate GDP  time series up to and including December 2019}. \gr{We then} 
forecast the next time stamp, \gr{January 2020,} to evaluate the model performance when excluding the Covid-19 pandemic. 
The relative errors of the MA forecasts per industry are represented as red points in Figure \ref{indrelerrcov}. To account for the reduced number of time stamps when excluding the pandemic, we scale the relative error by $\sqrt{T}$, where $T=48$ is the number of time stamps in our training  sample when excluding the Covid-19 period. 

Comparing the results to the period before Covid-19 \gr{validates o}ur intuition on the effect of Covid-19 on the forecasting performance,  
indicating the more challenging nature of the forecasting task for some industries under major disruptions in the economy. Nonetheless,  as indicated by the mean and sd of the relative error across releases, for most industries we are able to accurately forecast GDP values.
 
\begin{figure}[htb!]
    \centering
    \includegraphics[scale=.4]{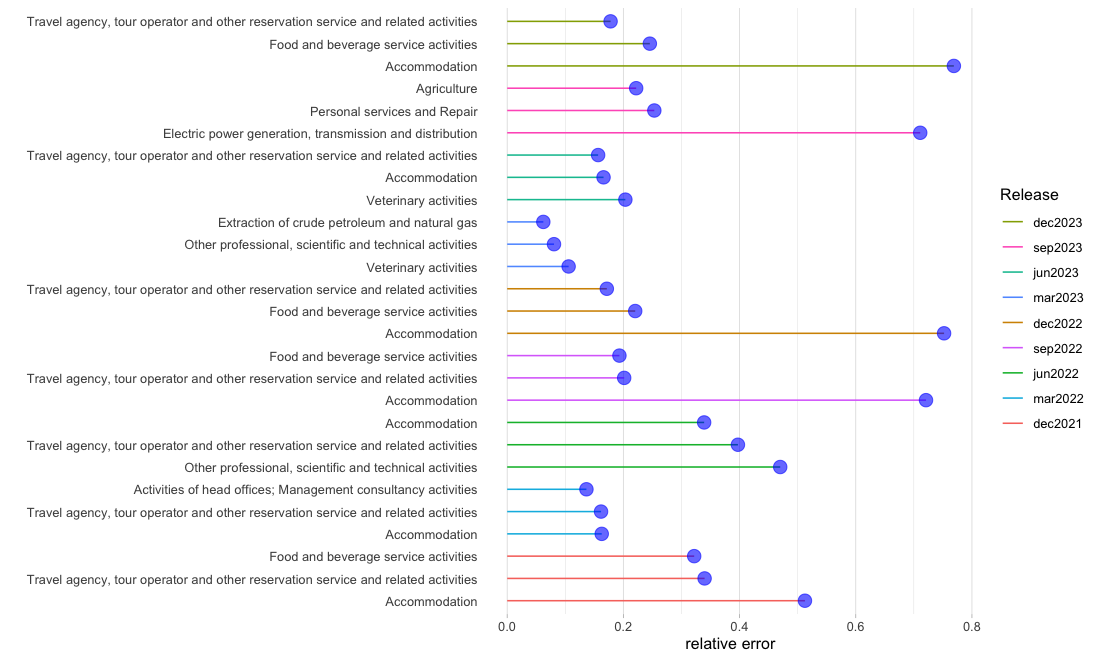}
    \caption{Top 3 industries (y axis) with highest relative error (x axis) after model averaging across lags$=1,\ldots,9$, for stage-1 neighbours, across 9 data releases (colour indicating \gr{release}).}
    \label{top3ind}
\end{figure}

Figure \ref{top3ind} \gr{shows}
the  
three industries with highest relative error,  
\gr{by} data release (\gr{with line colour indicating  release}). As expected from 
Figure \ref{indrelerrcov}, ``Accommodation'' and ``Travel Agency'' are the industries with highest relative error across releases. We note here that the reported forecasting performance is for time periods including the pandemic; here  we do not evaluate 
the forecasting performance of the model before the pandemic. We observe that \gr{the most challenging forecasting tasks are for ``Accommodation'', } 
for September and December releases,  which correspond to forecasting April and July time stamps. The ``Electric power generation, transmission and distribution'' industry appears as one of the top 3 industries with highest relative error only for the data release of September 2023, corresponding to GDP forecasting for April 2023.
This relatively \gr{poor} performance of the model 
may be related to differences in the time of recording of electricity-related payments and GDP accounting of the sector. Electricity contracts often are long-term, with regular payments and sluggish price adjustments; these payments would be recorded in the payment network time series. In contrast, national accounting rules applied to GDP aim to record the time and value of electricity generation and distribution at the time when the electricity enters the production process \citep{hotte2024}. 

\section{Conclusions}
\label{sec:conclusions}

This paper introduces a novel network approach for GDP nowcasting using payments data. The GNAR-ex model, as we call it, is a network autoregressive model tailored for networks with nodal time series and time-varying edge weights, that allows inferences of network effect parameters utilising the inherent network structure of the payments data. 

\gr{Using the GNAR-ex model for GDP nowcasting often yields more accurate results than models which do not include network information.}  
\gr{From an economics viewpoint there are  
more economic industry- and macroeconomic variables  which are commonly controlled for in economic nowcasting models and could be added in our GNAR-ex model application. Examining the value of adding further economic control variables \gr{and refining the nowcasts is work in progress.}}

While in this paper we applied the GNAR-ex model to GDP nowcasting, 
this model can also be implemented on similar data sets which can be represented using a network with time series on the nodes and edges.
\gr{There are a number of theoretical improvements possible. One could assume different parameters for the Gaussian errors for the nodal and the edge time series. It could also be interesting to consider non-Gaussian errors for some time series. Moreover, the current model assumes stationarity; there may be change points in the time series, such as around Covid-19 restrictions. Identifying change points is another area for future work.}
\mc{Yet another potential avenue to investigate pertains to building a factor model for the multivariate time series, which could further enhance the downstream forecasting task.} 

\gdr{The empirical work in this article relies on an experimental version of the payment data and quarterly published GVA data, as were publicly available when conducting the research. Improvements of the payment data by coverage and granularity are underway, and  coupled with higher GVA frequency data this  will provide 
a valuable basis for future work on GDP nowcasting based on ONS data.}

\medskip
\gr{{\bf Acknowledgements.} The authors acknowledge many helpful discussions with Andrew Walton, Andreina Naddeo, Keith Lai, Joseph Colliass, and  Dragos Cozma at the UK Office for National Statistics, and Alex Holmes, while at the UK Office for National Statistics.}

\begin{appendices}
\section{Appendix}
\label{sec:appendix}

\subsection{Additional results from simulation experiments for prediction performance of GNAR-ex}

\begin{figure}[htb!]
\centering
\vfill
    \includegraphics[scale=.45]{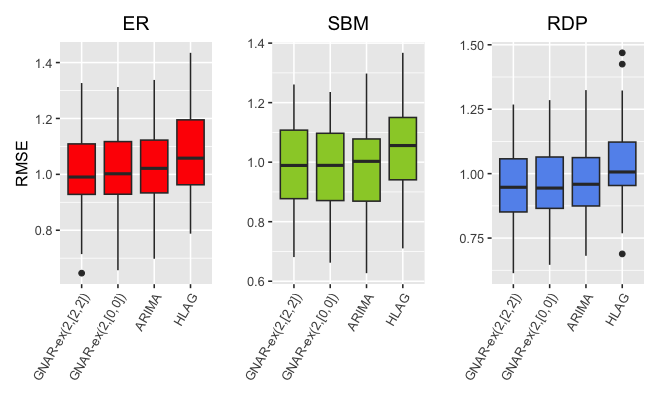}
    \caption{Prediction performance of GNAR-ex model with and without neighbourhood structure, ARIMA model and VAR model using the HLAG approach by \cite{nicholson2020high}, for nodal time series.}
    \label{nodeRMSEpred}
\end{figure}

Figure \ref{nodeRMSEpred} shows the prediction performance of the GNAR-ex model compared to an ARIMA and VAR model, for nodal time series only. We note that the results are based only on 10\% of the multivariate time series simulated in each repetition (20 nodal time series out of the approximately 180 time series simulated on edges and nodes combined).

\subsection{Additional results from simulation experiments from MA}

\begin{figure}[htb!]
    \centering
    \includegraphics[scale=.4]{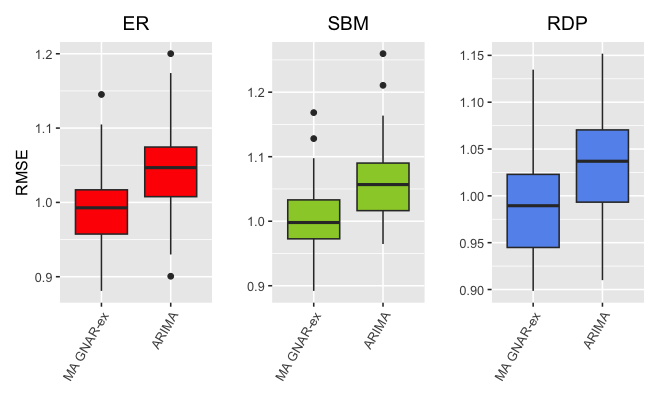}
    \caption{Prediction performance for all time series (nodal and edge) from MA across various lag sizes ($l=1,\ldots,9$) the GNAR-ex model with 1-stage neighbours compared to best fitting ARIMA model, for simulation regime 1 in Table \ref{simreglag1}.}
    \label{edgenodeRMSEpredMAreg1}
\end{figure}

\begin{figure}[htb!]
\centering
\vfill
    \includegraphics[scale=.4]{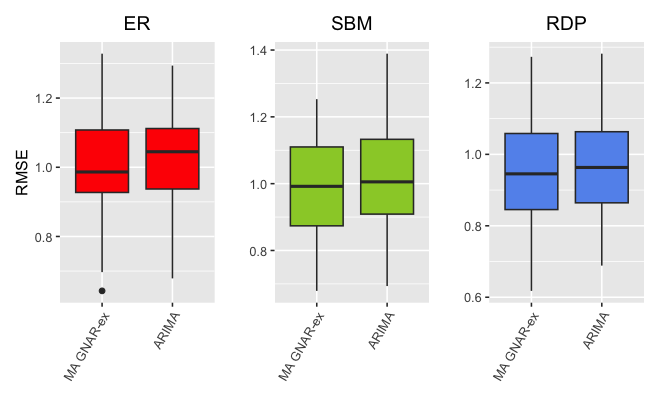}
    \caption{Prediction performance for nodal time series from MA across various lag sizes ($l=1,\ldots,9$) the GNAR-ex model with 1-stage neighbours compared to best fitting ARIMA model, for simulation regime 1 in Table \ref{simreglag1}.}
    \label{nodeRMSEpredMAreg1}
\end{figure}

Figures \ref{edgenodeRMSEpredMAreg1}, \ref{nodeRMSEpredMAreg1} and Table \ref{forecastunionreg1} show the results from implementing MA for the GNAR-ex model with 1-stage neighbour structure and various lag sizes ($l=1,\ldots,9$), for simulation regime 1 in Table \ref{simreglag1}.


\begin{table}[htb!]
\centering
\begin{tabular}{lllll}
             & \multicolumn{4}{l}{\textbf{Forecast inclusion}} \\
             & 17      & 18        & 19      & 20         \\ \hline
\textbf{ER}  & 0.14       & 0.14       & 0.42      & 0.3      \\
\textbf{SBM} & 0.04       & 0.2        & 0.48      & 0.28      \\
\textbf{RDP} & 0.06       & 0.08       & 0.38      & 0.48  
\end{tabular}
\caption{Proportion of repetitions for which  
\gr{exactly 17 (85\%), 18 (90\%),  19 (95\%) or 20 (100\%) of the 20} 
nodal time series (true last time stamp) lie within the union 95\% forecast interval, for different network structures (ER,SBM,RDP), for simulation regime 1 in Table \ref{simreglag1}.}
\label{forecastunionreg1}
\end{table}
\newpage
\subsection{GDP nowcasting: Forecast intervals}\label{sec:app_realdata}

In Tables \ref{predint_mar22}-\ref{predint_dec23}, we report the proportion of times that the true industry-level GDP values (node-level) lie within the 95\% forecast intervals obtained from the GNAR-ex model, for different releases,  
and for the different lag and stages.

\begin{table}[htb!]
\centering
\begin{tabular}{rrrrr}
  \hline
lag & stage 0 & stage 1 & stage 2 & stage 3 \\ 
  \hline
 1 & 0.97 & 1.00 & 0.97 & 0.97 \\ 
   2 & 0.97 & 1.00 & 0.94 & 0.97 \\ 
   3 & 0.94 & 0.94 & 0.97 & 0.97 \\ 
   4 & 0.97 & 0.94 & 0.94 & 0.94 \\ 
   5 & 0.97 & 0.94 & 0.97 & 0.94 \\ 
   6 & 0.94 & 0.97 & 0.97 & 0.97 \\ 
   7 & 0.88 & 0.84 & 0.84 & 0.88 \\ 
   8 & 0.91 & 0.91 & 0.78 & 0.91 \\ 
   9 & 0.88 & 0.91 & 1.00 & 0.91 \\ 
   \hline
\end{tabular}
\caption{Number of times that true industry level GDP values (original scale) lie within the 95\% forecast intervals, among industries, for each lag and stages of neighbours, for \gr{the} data release of March 2022.}
\label{predint_mar22}
\end{table}

\begin{table}[htb!]
\centering
\begin{tabular}{rrrrr}
  \hline
lag & stage 0 & stage 1 & stage 2 & stage 3 \\ 
  \hline
1 & 0.75 & 0.75 & 0.75 & 0.75 \\ 
  2 & 0.72 & 0.72 & 0.69 & 0.69 \\ 
  3 & 0.69 & 0.69 & 0.72 & 0.72 \\ 
  4 & 0.75 & 0.72 & 0.69 & 0.69 \\ 
  5 & 0.75 & 0.72 & 0.62 & 0.66 \\ 
  6 & 0.75 & 0.69 & 0.69 & 0.75 \\ 
  7 & 0.72 & 0.62 & 0.72 & 0.69 \\ 
  8 & 0.75 & 0.62 & 0.69 & 0.69 \\ 
  9 & 0.72 & 0.66 & 0.31 & 0.59 \\ 
   \hline
\end{tabular}
\caption{Number of times that true industry level GDP values (original scale) lie within the 95\% forecast intervals, among industries, for each lag and stages of neighbours, for \gr{the} data release of June 2022.}
\label{predint_jun22}
\end{table}

\begin{table}[htb!]
\centering
\begin{tabular}{rrrrr}
  \hline
lag & stage 0 & stage 1 & stage 2 & stage 3 \\ 
  \hline
1 & 1.00 & 1.00 & 1.00 & 1.00 \\ 
  2 & 1.00 & 1.00 & 0.97 & 0.97 \\ 
  3 & 1.00 & 1.00 & 0.97 & 0.97 \\ 
  4 & 1.00 & 1.00 & 1.00 & 0.94 \\ 
  5 & 1.00 & 1.00 & 1.00 & 1.00 \\ 
  6 & 1.00 & 0.97 & 1.00 & 1.00 \\ 
  7 & 0.97 & 1.00 & 0.91 & 0.88 \\ 
  8 & 1.00 & 0.97 & 0.94 & 0.91 \\ 
  9 & 0.97 & 1.00 & 0.91 & 0.91 \\ 
   \hline
\end{tabular}
\caption{Number of times that true industry level GDP values (original scale) lie within the 95\% forecast intervals, among industries, for each lag and stages of neighbours, for \gr{the} data release of September 2022.}
\label{predint_sep22}
\end{table}

\begin{table}[htb!]
\centering
\begin{tabular}{rrrrr}
  \hline
lag & stage 0 & stage 1 & stage 2 & stage 3 \\ 
  \hline
1 & 1.00 & 1.00 & 1.00 & 1.00 \\ 
  2 & 1.00 & 1.00 & 1.00 & 1.00 \\ 
  3 & 1.00 & 1.00 & 1.00 & 1.00 \\ 
  4 & 1.00 & 1.00 & 1.00 & 1.00 \\ 
  5 & 1.00 & 1.00 & 1.00 & 1.00 \\ 
  6 & 1.00 & 1.00 & 1.00 & 1.00 \\ 
  7 & 0.97 & 1.00 & 0.94 & 0.91 \\ 
  8 & 1.00 & 0.97 & 1.00 & 0.94 \\ 
  9 & 0.97 & 0.97 & 0.94 & 0.94 \\ 
   \hline
\end{tabular}
\caption{Number of times that true industry level GDP values (original scale) lie within the 95\% forecast intervals, among industries, for each lag and stages of neighbours, for \gr{the} data release of December 2022.}
\label{predint_dec22}
\end{table}

\begin{table}[htb!]
\centering
\begin{tabular}{rrrrr}
  \hline
lag & stage 0 & stage 1 & stage 2 & stage 3 \\ 
  \hline
1 & 1.00 & 1.00 & 1.00 & 1.00 \\ 
  2 & 1.00 & 1.00 & 1.00 & 1.00 \\ 
  3 & 1.00 & 1.00 & 1.00 & 1.00 \\ 
  4 & 1.00 & 1.00 & 1.00 & 1.00 \\ 
  5 & 1.00 & 1.00 & 1.00 & 1.00 \\ 
  6 & 1.00 & 1.00 & 1.00 & 1.00 \\ 
  7 & 1.00 & 1.00 & 1.00 & 1.00 \\ 
  8 & 1.00 & 1.00 & 1.00 & 1.00 \\ 
  9 & 1.00 & 1.00 & 1.00 & 1.00 \\ 
   \hline
\end{tabular}
\caption{Number of times that true industry level GDP values (original scale) lie within the 95\% forecast intervals, among industries, for each lag and stages of neighbours, for \gr{the} data release of March 2023.}
\label{predint_mar23}
\end{table}
\clearpage
\newpage
\begin{table}[htb!]
\centering
\begin{tabular}{rrrrr}
  \hline
lag & stage 0 & stage 1 & stage 2 & stage 3 \\ 
  \hline
1 & 0.97 & 1.00 & 1.00 & 1.00 \\ 
  2 & 0.94 & 0.94 & 0.66 & 0.66 \\ 
  3 & 0.81 & 0.66 & 0.62 & 0.62 \\ 
  4 & 0.91 & 0.84 & 0.69 & 0.88 \\ 
  5 & 0.88 & 0.75 & 1.00 & 0.75 \\ 
  6 & 0.84 & 0.72 & 0.69 & 0.66 \\ 
  7 & 0.78 & 0.66 & 0.44 & 0.44 \\ 
  8 & 0.75 & 0.66 & 0.44 & 0.47 \\ 
  9 & 0.72 & 0.56 & 0.59 & 0.53 \\ 
   \hline
\end{tabular}
\caption{Number of times that true industry level GDP values (original scale) lie within the 95\% forecast intervals, among industries, for each lag and stages of neighbours, for \gr{the} data release of June 2023.}
\label{predint_jun23}
\end{table}

\begin{table}[htb!]
\centering
\begin{tabular}{rrrrr}
  \hline
lag & stage 0 & stage 1 & stage 2 & stage 3 \\ 
  \hline
1 & 0.66 & 0.75 & 0.62 & 0.62 \\ 
  2 & 0.66 & 0.72 & 0.66 & 0.66 \\ 
  3 & 0.72 & 0.69 & 0.66 & 0.66 \\ 
  4 & 0.62 & 0.66 & 0.62 & 0.62 \\ 
  5 & 0.69 & 0.66 & 0.50 & 0.66 \\ 
  6 & 0.69 & 0.66 & 0.62 & 0.66 \\ 
  7 & 0.66 & 0.66 & 0.62 & 0.62 \\ 
  8 & 0.59 & 0.59 & 0.66 & 0.62 \\ 
  9 & 0.59 & 0.59 & 0.66 & 0.66 \\ 
   \hline
\end{tabular}
\caption{Number of times that true industry level GDP values (original scale) lie within the 95\% forecast intervals, among industries, for each lag and stages of neighbours, for \gr{the} data release of September 2023.}
\label{predint_sep23}
\end{table}

\begin{table}[htb!]
\centering
\begin{tabular}{rrrrr}
  \hline
lag & stage 0 & stage 1 & stage 2 & stage 3 \\ 
  \hline
1 & 1.00 & 1.00 & 1.00 & 1.00 \\ 
  2 & 1.00 & 1.00 & 1.00 & 1.00 \\ 
  3 & 1.00 & 1.00 & 1.00 & 1.00 \\ 
  4 & 1.00 & 1.00 & 1.00 & 0.81 \\ 
  5 & 1.00 & 0.97 & 1.00 & 1.00 \\ 
  6 & 1.00 & 0.97 & 1.00 & 1.00 \\ 
  7 & 1.00 & 0.97 & 1.00 & 1.00 \\ 
  8 & 1.00 & 0.97 & 0.97 & 0.97 \\ 
  9 & 0.97 & 0.97 & 1.00 & 0.97 \\ 
   \hline
\end{tabular}
\caption{Number of times that true industry level GDP values (original scale) lie within the 95\% forecast intervals, among industries, for each lag and stages of neighbours, for \gr{the} data release of December 2023.}
\label{predint_dec23}
\end{table}

\end{appendices}

\section{Disclaimer}
The views expressed are those of the authors and may not reflect the views of the Office for National Statistics or the wider UK Government.

\section{Competing interests}
No competing interest is declared.

\section{Funding}
The authors gratefully acknowledge the ONS-Turing Strategic partnership for funding. G.R. also
acknowledges funding from the Engineering and Physical Sciences Research Council (EPSRC) grants
EP/T018445/1, EP/R018472/1,  and EP/X0021951.

\section{Author contributions statement}

\gr{A.M.,  G.R. and M.C. conceived the model. A.M. carried out most of the experiments. K.H. processed the data and provided the economics context as well as some exploratory data analysis. A.M. and K.H. drafted the manuscript. All authors analysed the results and reviewed the manuscript.}

\bibliography{sample}

\end{document}